\documentstyle[11pt]{article}
\newcommand{\be}{\begin{equation}}
\newcommand{\ee}{\end{equation}}
\begin{document}
\pagestyle{plain}
\huge
\title{\bf Synchrotron production of photons by the two-body system}
\large
\author{\bf Miroslav Pardy\\
Department of Theoretical Physics and Astrophysics\\
Faculty of Science, Masaryk University\\
Kotl\'{a}\v{r}sk\'{a} 2, 611 37 Brno, Czech Republic\\
E-mail:pamir@physics.muni.cz}
\date{}
\maketitle
\vspace {30mm}
\large
\begin{abstract}
\large
The power spectrum formula of the synchrotron radiation generated by the
motion of a two-body charged system in an accelerator, is derived
in the framework of the Schwinger source theory. The final formula can be
used to verify the Lorentz length contraction of the two-body
system moving in the synchrotron. 
\end{abstract}
\vspace{50mm}

\newpage

\hspace{3ex}
   The production of photons by circular motion of charged particle
in an accelerator is one of the most interesting problems in the classical
and quantum electrodynamics.

In this paper we are interested in the photon production
initiated by circular motion and a two-body charged system.
This process specifies the synergic synchrotron \v{C}erenkov
radiation, which was calculated in source theory two decades ago
by Schwinger et al. (1976). We will follow also the the article
Pardy, (1994a).
The synergic process includes the effect of the medium,
which is represented by the
phenomenological index of refraction $n$, and it is well known that this
phenomenological constant depends on the external magnetic field.

We will investigate, first, how the original Schwinger et al. spectral
formula of the synergic synchrotron \v{C}erenkov radiation of a
charged particle moving in a medium is modified if
we consider a two-body system.
Then, we will treat this process in vacuum.
This problem is an analogue of the linear problem solved recently
by the author (Pardy, 1997) also in source theory. We will show
that the original spectral formula
of the synergic synchrotron-\v Cerenkov radiation is modulated by function
$\cos^{2}(a\omega/2v)$, where $a$ is the distance between charges, $v$ is
their velocity and $\omega$ is the frequency of the synergic radiation
produced by the system.

Source theory (Schwinger, 1970, 1973; Dittrich, 1978)
 was initially constructed
for a description of high-energy particle physics experiments.
It was found that the
original formulation simplifies the calculations in
electrodynamics and gravity, where the interactions are mediated by
the photon and graviton, respectively. It simplifies particularly the
calculations with radiative corrections (Dittrich, 1978; Pardy, 1994b) .

The basic formula of the Schwinger nsource theory is the so called
vacuum to vacuum amplitude:
\be
\label{1}
\langle 0_{+}|0_{-} \rangle = e^{\frac{i}{\hbar}\*W},
\ee
where, for the case of the electromagnetic field in the medium, the action
$W$ is given by
\be
\label{2}
W = \frac{1}{2c^2}\*\int\,(dx)(dx')J^{\mu}(x){D}_{+\mu\nu}(x-x')J^{\nu}(x'),
\ee
where
\be
\label{3}
{D}_{+}^{\mu\nu} = \frac{\mu}{c}[g^{\mu\nu} +
(1-n^{-2})\beta^{\mu}\beta^{\nu}]\*{D}_{+}(x-x'),
\ee
and $\beta^{\mu}\, \equiv \, (1,{\bf 0})$, $J^{\mu}\, \equiv \,(c\varrho,{\bf
J})$ is the conserved current, $\mu$ is the magnetic permeability of
the medium, $\epsilon$ is the dielectric constant od the medium, and
$n=\sqrt{\epsilon\mu}$ is the index of refraction of the medium.
Function ${D}_{+}$ is defined as follows (Schwinger et al., 1976):

\be
\label{4}
D_{+}(x-x') =\frac {i}{4\pi^2\*c}\*\int_{0}^{\infty}d\omega
\frac {\sin\frac{n\omega}{c}|{\bf x}-{\bf x}'|}{|{\bf x} - {\bf x}'|}\*
e^{-i\omega|t-t'|}.
\ee

The probability of the persistence of vacuum follows from the vacuum
amplitude (1) in the following form:

\be
\label{5}
|\langle 0_{+}|0_{-} \rangle|^2 = e^{-\frac{2}{\hbar}\*\rm Im\*W},
\ee
where ${\rm Im}\;W$ is the basis for the definition of the spectral
function $P(\omega,t)$ as follows:
\be
\label{6}
-\frac{2}{\hbar}\*{\rm Im}\;W \;\stackrel{d}{=} \; -\,
\int\,dtd\omega\frac{P(\omega,t)}
{\hbar\omega}.
\ee

Now, if we insert eq. (2) into eq. (6), we get,
after extracting $P(\omega,t)$, the following general expression
for this spectral function:

$$P(\omega,t) = -\frac{\omega}{4\pi^2}\*\frac{\mu}{n^2}\*\int\,d{\bf x}
d{\bf x}'dt'\left[\frac{\sin\frac{n\omega}{c}|{\bf x} -
{\bf x}'|}{|{\bf x} - {\bf x}'|}\right]\;\times $$

\be
\label{7}
\cos[\omega\*(t-t')]\*[\varrho({\bf x},t)\varrho({\bf x}',t')
- \frac{n^2}{c^2}\*{\bf J}({\bf x},t)\cdot{\bf J}({\bf x}',t')].
\ee

Now, we will apply the formula (7) to the two-body system
with the same charged
particles in order to get its synergic synchrotron-\v Cerenkov
radiation. The synchrotron
radiation is produced by particle of charge $e$
moving in a uniform circular motion with velocity ${\bf v}$
in the plane perpendicular to the
direction of the constant magnetic field ${\bf H}$ (chosen to be
in the $+z$ direction).

On the other hand the \v Cerenkov electromagnetic radiation is
generated by a fast-moving charged particle
in a medium when its speed is faster than the speed of light in this medium.
This radiation was first observed experimentally by
\v{C}erenkov (1936) and theoretically interpreted by
Tamm and Frank (1937) in
the framework of classical electrodynamics. A source
theoretical description of this effect was given by Schwinger et al.
(1976) in the zero-temperature regime, and the classical spectral
formula was generalized to the  finite-temperature situation in
electrodynamics and gravity in the
framework of the source theory by Pardy (1989, 1995).

In electrodynamics one usually considers synchrotron  radiation
produced by a uniformly moving charge
with constant orbital velocity. Here we consider the system of
two equal charges
$e$ with the constant mutual distance $a$
moving with orbital velocity $v$ in the accelerator.
For the sake of generality
we consider also that a dielectric medium is present.
So we write for the charge density $\varrho$ and for the current
density ${\bf J}$
of the two-body system:

\be
\label{8}
\varrho({\bf x},t) = e\*\delta\*({\bf x}-{\bf x_{1}}(t))  +
e\*\delta\*({\bf x}-{\bf x_{2}}(t))
\ee
and

\be
\label{9}
{\bf J}({\bf x},t) = e\*{\bf v}_{1}(t)\*\delta\*({\bf x}-{\bf x_{1}}(t)) +
e\*{\bf v}_{2}(t)\*\delta\*({\bf x}-{\bf x_{2}}(t))
\ee
with
\be
\label{10}
{\bf x}_{1}(t)  = {\bf x}(t) =
R({\bf i}\cos(\omega_{0}t) + {\bf j}\sin(\omega_{0}t)),
\ee

\be
\label{11}
{\bf x}_{2}(t) =
R({\bf i}\cos(\omega_{0}t + \delta\varphi) +
{\bf j}\sin(\omega_{0}t + \delta\varphi)) =
{\bf x}(t + \frac {\delta\varphi}{\omega_{0}});\quad
\delta\varphi = \frac {a}{R}.
\ee

We will suppose for simplicity that the distance of the particles forming the
two-body system is very small in comparison with the diameter $R$ of the
circle accelerator, which means that the velocities of both particles are
approximately the same, or ${\bf v}_{1}(t) \approx {\bf v}_{2}(t) =
{\bf v}(t)$, where ($H = |{\bf H}|, E =$ energy of a particle)
\be
\label{12}
{\bf v}(t) = d{\bf x}/dt, \hspace{5mm} \omega_{0} = v/R, \hspace{5mm}
R = \frac {\beta\*E}{eH}, \hspace{5mm}
\beta = v/c, \hspace{5mm} v = |{\bf v}|.
\ee

After insertion of eqs. (8) and (9) into eq. (7), and after some mathematical
operations we get

$$P(\omega,t) =
-\frac{\omega}{4\pi^2}\*\frac{\mu}{n^2}e^{2}\*\int_{-\infty}^{\infty}\,
dt'\cos(t-t')
\left[1 - \frac {{\bf v}(t)\cdot {\bf v}(t')}{c^{2}}n^{2}\right]
\;\times $$

$$\left\{\frac{\sin\frac {n\omega}{c}|{\bf x}_{1}(t) -{\bf x}_{1}(t')|}
{|{\bf x}_{1}(t) -{\bf x}_{1}(t')|} +
\frac{\sin\frac {n\omega}{c}|{\bf x}_{1}(t) -{\bf x}_{2}(t')|}
{|{\bf x}_{1}(t) -{\bf x}_{2}(t')|} + \right.$$

\be
\label{13}
\left.\frac{\sin\frac {n\omega}{c}|{\bf x}_{2}(t) -{\bf x}_{1}(t')|}
{|{\bf x}_{2}(t) -{\bf x}_{1}(t')|} +
\frac{\sin\frac {n\omega}{c}|{\bf x}_{2}(t) -{\bf x}_{2}(t')|}
{|{\bf x}_{2}(t) -{\bf x}_{2}(t')|}\right\}.
\ee

Using $t' = t + \tau$, we get
\be
\label{14}
{\bf x}_{1}(t) -{\bf x}_{1}(t') =  {\bf x}(t) -{\bf x}(t+\tau)
\stackrel{d}{=} {\bf A},
\ee

\be
\label{15}
{\bf x}_{1}(t) -{\bf x}_{2}(t') =  {\bf x}(t) -{\bf x}(t+\tau + \frac
{\delta\varphi}{\omega_{0}}) \stackrel{d}{=} {\bf B},
\ee

\be
\label{16}
{\bf x}_{2}(t) -{\bf x}_{1}(t') =
{\bf x}(t+\frac {\delta\varphi}{\omega_{0}})
-{\bf x}(t+\tau) \stackrel{d}{=} {\bf C},
\ee

\be
\label{17}
{\bf x}_{2}(t) -{\bf x}_{2}(t') =
{\bf x}(t+\frac {\delta\varphi}{\omega_{0}})
-{\bf x}(t+\tau + \frac {\delta\varphi}{\omega_{0}})
\stackrel{d}{=} {\bf D}.
\ee

Using geometrical representation of vectors ${\bf x}_{i}(t)$, we get

\be
\label{18}
|{\bf A}| = [R^{2} + R^{2} - 2RR\cos(\omega_{0}\tau)]^{1/2} =
2R\left|\sin\left(\frac {\omega_{0}\tau}{2}\right)\right|,
\ee

\be
\label{19}
|{\bf B}| =  2R\left|\sin\left(\frac {\omega_{0}\tau +
\delta\varphi}{2}\right)\right|,
\ee

\be
\label{20}
|{\bf C}| =  2R\left|\sin\left(\frac {\omega_{0}\tau -
\delta\varphi}{2}\right)\right|,
\ee

\be
\label{21}
|{\bf D}| =  2R\left|\sin\left(\frac {\omega_{0}\tau}{2}\right)\right|.
\ee

Using

\be
\label{22}
{\bf v}(t)\cdot{}{\bf v}(t+\tau) = \omega^{2}_{0}R^{2}\cos\omega_{0}\tau,
\ee
and relations (18)--(21), we get, with $v= \omega_{0}R$

$$P(\omega,t) =
-\frac{\omega}{4\pi^2}\*\frac{\mu}{n^2}e^{2}\*\int_{-\infty}^{\infty}\,
d\tau \cos\omega\tau
\left[1 - \frac {n^{2}}{c^{2}}v^{2}\cos\omega_{0}\tau\right]
\;\times $$

$$\left\{\frac{\sin\left[\frac {2Rn\omega}{c}\sin
\left(\frac {\omega_{0}\tau}{2}\right)\right]}
{2R\sin\left(\frac {\omega_{0}\tau}{2}\right)} +
\frac{\sin\left[\frac {2Rn\omega}{c}
\sin\left(\frac {\omega_{0}\tau + \delta\varphi}
{2}\right)\right]}
{2R\sin\left(\frac {\omega_{0}\tau+\delta\varphi}{2}\right)}\quad +
\right.$$

\be
\label{23}
\left.\frac{\sin\left[\frac {2Rn\omega}{c}\sin
\left(\frac {\omega_{0}\tau-\delta\varphi}{2}\right)\right]}
{2R\sin\left(\frac {\omega_{0}\tau-\delta\varphi}{2}\right)} +
\frac{\sin\left[\frac {2Rn\omega}{c}
\sin\left(\frac {\omega_{0}\tau}
{2}\right)\right]}
{2R\sin\left(\frac {\omega_{0}\tau}{2}\right)}\right\}.
\ee

Introducing new variable $T$ by relation

\be
\label{24}
\omega_{0}\tau + \alpha_{i} = \omega_{0}T
\ee
for every integral in Eq. (23), where

\be
\label{25}
\alpha_{i} =  0,\quad \delta\varphi, \quad -\delta\varphi, \quad 0,
\ee
we get $P(\omega,t)$ in the following form

$$P(\omega,t) =
-\frac{\omega}{4\pi^2}\frac {e^{2}}{2R}
\*\frac{\mu}{n^2}\*\int_{-\infty}^{\infty} dT \sum_{i=1}^{4}\times $$

\be
\label{26}
\cos(\omega T - \frac {\omega}{\omega_{0}}\alpha_{i})
\left[1 - \frac {c^{2}}{n^{2}}v^{2}\cos(\omega_{0} T -
\alpha_{i})\right]
\left\{\frac{\sin\left[\frac {2Rn\omega}{c}\sin
\left(\frac {\omega_{0}T}{2}\right)\right]}
{\sin\left(\frac {\omega_{0}T}{2}\right)}\right\}.
\ee
The last formula can be written in the more compact form,

\be
\label{27}
P(\omega,t) = -\frac {\omega}{4\pi^{2}}\frac {\mu}{n^{2}}\frac {e^{2}}{2R}
\sum_{i=1}^{4}\left\{P_{1}^{(i)} -\frac {n^{2}}{c^{2}}v^{2}
P_{2}^{(i)}\right\},
\ee
where

\be
\label{28}
P_{1}^{(i)} = J_{1a}^{(i)}\cos\frac {\omega}{\omega_{0}}\alpha_{i} +
J_{1b}^{(i)}\sin\frac {\omega}{\omega_{0}}\alpha_{i}
\ee
and

$$P_{2}^{(i)} = J_{2A}^{(i)}\cos\alpha_{i}
\cos\frac {\omega}{\omega_{0}}\alpha_{i} +$$

\be
\label{29}
J_{2B}^{(i)}\cos\alpha_{i}\sin\frac {\omega}{\omega_{0}}\alpha_{i} +
J_{2C}^{(i)}\sin\alpha_{i}\cos\frac {\omega}{\omega_{0}}\alpha_{i} +
J_{2D}^{(i)}\sin\alpha_{i}\sin\frac {\omega}{\omega_{0}}\alpha_{i},
\ee
where

\be
\label{30}
J_{1a}^{(i)} = \int_{-\infty}^{\infty}dT\cos\omega T
\left\{\frac{\sin\left[\frac {2Rn\omega}{c}\sin
\left(\frac {\omega_{0}T}{2}\right)\right]}
{\sin\left(\frac {\omega_{0}T}{2}\right)}\right\},
\ee

\be
\label{31}
J_{1b}^{(i)} = \int_{-\infty}^{\infty}dT\sin\omega T
\left\{\frac{\sin\left[\frac {2Rn\omega}{c}\sin
\left(\frac {\omega_{0}T}{2}\right)\right]}
{\sin\left(\frac {\omega_{0}T}{2}\right)}\right\},
\ee

\be
\label{32}
J_{2A}^{(i)} = \int_{-\infty}^{\infty}dT\cos\omega_{0}T\cos\omega T
\left\{\frac{\sin\left[\frac {2Rn\omega}{c}\sin
\left(\frac {\omega_{0}T}{2}\right)\right]}
{\sin\left(\frac {\omega_{0}T}{2}\right)}\right\},
\ee

\be
\label{33}
J_{2B}^{(i)} = \int_{-\infty}^{\infty}dT\cos\omega_{0}T\sin\omega T
\left\{\frac{\sin\left[\frac {2Rn\omega}{c}\sin
\left(\frac {\omega_{0}T}{2}\right)\right]}
{\sin\left(\frac {\omega_{0}T}{2}\right)}\right\},
\ee

\be
\label{34}
J_{2C}^{(i)} = \int_{-\infty}^{\infty}dT\sin\omega_{0}T\cos\omega T
\left\{\frac{\sin\left[\frac {2Rn\omega}{c}\sin
\left(\frac {\omega_{0}T}{2}\right)\right]}
{\sin\left(\frac {\omega_{0}T}{2}\right)}\right\},
\ee

\be
\label{35}
J_{2D}^{(i)} = \int_{-\infty}^{\infty}dT\sin\omega_{0}T\sin\omega T
\left\{\frac{\sin\left[\frac {2Rn\omega}{c}\sin
\left(\frac {\omega_{0}T}{2}\right)\right]}
{\sin\left(\frac {\omega_{0}T}{2}\right)}\right\}.
\ee
Using

\be
\label{36}
\omega_{0}T = \varphi + 2\pi\*l,  \hspace{7mm} \varphi\in(-\pi,\pi),\;
l = 0,\, \pm1,\, \pm2,\, ... ,
\ee
we can transform the $T$-integral into the sum of the telescopic
integrals according to the scheme

\be
\label{37}
\int_{-\infty}^{\infty}dT\quad\longrightarrow \quad\frac {1}{\omega_{0}}
\sum_{-\infty}^{\infty}\int_{-\pi}^{\pi}d\varphi.
\ee

Using the fact that for the odd functions $f(\varphi)$ and $g(l)$,
the relations are valid,

\be
\label{38}
\int_{-\pi}^{\pi}f(\varphi)d\varphi = 0; \quad \sum_{l=-\infty}^{\infty}g(l)
= 0,
\ee
we can write

\be
\label{39}
J_{1a}^{(i)} = \frac {1}{\omega_{0}}\sum_{l}\int_{-\pi}^{\pi}
d\varphi\left\{\cos{\frac {\omega}{\omega_{0}}\varphi\cos{2\pi l}
\frac{\omega}{\omega_{0}}}\right\}
\left\{\frac{\sin\left[\frac {2Rn\omega}{c}\sin
\left(\frac {\varphi}{2}\right)\right]}
{\sin\left(\frac {\varphi}{2}\right)}\right\},
\ee

\be
\label{40}
J_{1b}^{(i)} = 0.
\ee

For integrals with indices A,B,C,D we get:

\be
\label{41}
J_{2A}^{(i)} = \frac {1}{\omega_{0}}\sum_{l}\int_{-\pi}^{\pi}
d\varphi\cos\varphi
\left\{\cos{\frac {\omega}{\omega_{0}}\varphi\cos{2\pi l}
\frac{\omega}{\omega_{0}}}\right\}
\left\{\frac{\sin\left[\frac {2Rn\omega}{c}\sin
\left(\frac {\varphi}{2}\right)\right]}
{\sin\left(\frac {\varphi}{2}\right)}\right\},
\ee

\be
\label{42}
J_{2B}^{(i)} = J_{2C}^{(i)} = 0,
\ee
\be
\label{43}
J_{2D}^{(i)} = \frac {1}{\omega_{0}}\sum_{l}\int_{-\pi}^{\pi}
d\varphi\sin\varphi
\left\{\sin{\frac {\omega}{\omega_{0}}\varphi\cos 2\pi l
\frac{\omega}{\omega_{0}}}\right\}
\left\{\frac{\sin\left[\frac {2Rn\omega}{c}\sin
\left(\frac {\varphi}{2}\right)\right]}
{\sin\left(\frac {\varphi}{2}\right)}\right\}.
\ee

Using the Poisson theorem,

\be
\label{44}
\sum_{k=-\infty}^{\infty}\cos 2\pi\frac {\omega}{\omega_{0}}k  =
\sum_{k=-\infty}^{\infty}\omega_{0}\delta(\omega - \omega_{0}k),
\ee
the definition of the Bessel functions $J_{2l}$, and their
corresponding derivation and integral

\be
\label{45}
\frac {1}{2\pi}\int_{-\pi}^{\pi}d\varphi\cos\left(z\sin\frac {\varphi}{2}
\right)\cos l\varphi  = J_{2l}(z),
\ee

\be
\label{46}
\frac {1}{2\pi}\int_{-\pi}^{\pi}d\varphi\sin\left(z\sin\frac {\varphi}{2}
\right)\cos l\varphi  = - J'_{2l}(z),
\ee

\be
\label{47}
\frac {1}{2\pi}\int_{-\pi}^{\pi}d\varphi
\frac{\sin\left(z\sin\frac{\varphi}{2}\right)}
{\sin(\varphi/2)}\cos l\varphi  = \int_{0}^{z}J_{2l}(x)dx,
\ee
we get, with ($a \ll R$)

\be
\label{48}
\sin\alpha_{i}\sin\frac {\omega}{\omega_{0}}\alpha_{i} \approx 0
\ee
and with the definition of the partial power spectrum $P_{l}$,

\be
\label{49}
P(\omega) = \sum_{l=1}^{\infty}  \delta(\omega - l\omega_{0})P_{l},
\ee
the following final form of the partial power spectrum generated by motion of
two-charge system moving in the cyclotron:

\be
\label{50}
P_{l}(\omega,t) = \cos^{2}\left(\frac {a\omega}{2v}
\right)\frac {e^2}{\pi\*n^2}\*\frac {\omega\mu\omega_{0}}{v}\*
\left(2n^2\beta^2J'_{2l}(2ln\beta) -
(1 - n^2\*\beta^2)\*\int_{0}^{2ln\beta}dxJ_{2l}(x)\right).
\ee

Our goal is to apply the last formula to the situation
in medium of the accelerator where it is in fact, a vacuum. In
this case we can put $\mu = 1, n = 1$ in the last formula and so we have

\be
\label{51}
P_{l} =
\cos^{2}\left(\frac {a\omega}{2v}\right)
\frac {e^2}{\pi}\*\frac {\omega\omega_{0}}{v}\*
\left(2\beta^2J'_{2l}(2l\beta) -
(1 - \beta^2)\*\int_{0}^{2l\beta}dxJ_{2l}(x)\right).
\ee

Using the approximative formulas

\be
\label{52}
J'_{2l}(2l\beta) \sim \frac{1}{\sqrt{3}}\*\frac {1}{\pi}\*
\left(\frac {3}{2l_{c}}\right)^{2/3}\*K_{2/3}(l/l_{c}),
\quad  l \gg1,
\ee

\be
\label{53}
\int_{0}^{2l\beta}J_{2l}(y)dy  \sim \frac{1}{\sqrt{3}}\*\frac {1}{\pi}\*
\int_{l/l_{c}}^{\infty}K_{1/3}(y)dy,
\quad l \gg1,
\ee
with (Schwinger et al., 1976)

\be
\label{54}
l_c = \frac {3}{2}(1-\beta^2)^{-3/2},
\ee
substituting Eqs. (52) and (53) into Eq. (51),
respecting the high-energy situation for the
high-energy particles where $(1-\beta^2) \rightarrow 0$,
and using the recurence relation

\be
\label{55}
K'_{2/3} = -\frac{1}{2}(K_{1/3} + K_{5/3}),
\ee
and definition function $\kappa(\xi)$
\be
\label{56}
\kappa(\xi) = \xi\*\int_{\xi}^{\infty}K_{5/3}(y)dy , \quad
\xi = l/l_{c},
\ee
we get power spectrum formula of electron-electron pair as follows:

$$P(\omega) = \cos^{2}\left(\frac {a\omega}{2v}\right)
\frac{\omega\*e^2}{\pi^{2}\*R}\*\sqrt{\frac {\pi}{6}}
\*\left(\frac {3}{2l}\right)^{2/3}\xi^{1/6}\*e^{-\xi};\quad
l = \frac {\omega}{\omega_{0}}. \eqno(57)$$
where we used the idea that the discrete spectrum parametrized by number
$l$ is effectively continuous for $l\gg 1$. In such a case there is
an relation

$$P(\omega) = P_{(l = \omega/\omega_{0})}\*\left(\frac {1}{\omega_{0}}\right).
\eqno(58)$$

Formula (57) is analogous to the formula derived in Pardy (1997)
 for the
linear motion of a two-charge system emitting \v Cerenkov radiation.

The radiative corrections
obviously influence the spectrum (Schwinger, 1970 and
Pardy, 1994b). Determination of this phenomenon forms the special
problem of the accelerator physics.

Use of large accelerators,for instance, Grenoble, DESY, CERN should make
possible experimental verification of the derived
formulas involving also the Lorentz contraction.
Instead of two electrons
we can consider, say,  two bunches with 10$^{10}$ electrons in each
bunch of volume 300$ \mu$m $\times$ 40$ \mu$m $\times$ 0.01 m, with the
rest distance $l = 1$ m between them.
The distance between the bunches is the
relativistic length $a$ and it can be determined by the
synchrotron spectrum derived here.

The results of the Lorentz contraction measurement obtained
from the synchrotron radiation spectrum in vacuum  ($n = 1, \mu = 1$)
should be
identical with the results of a measurement obtained from a spectrum generated
by linear uniform particle motion in a medium because the interference
of light in vacuum does not differ from the interference
of light emitted by the \v Cerenkov effect in a medium.

\vspace{10mm}

\begin{center}
{\bf References}
\end{center}
\vspace{10mm}
\noindent
\v{C}erenkov, P. A., C.R., Acad. Sci. (USSR) {\bf 3} (1936), 413.\\
Dittrich, W., Fortschritte der Physik, {\bf 26} (1978), 289.\\
Pardy, M., Phys. Lett. A {\bf 134} (1989), 357.\\
Pardy, M.,  Phys. Lett. A {\bf 189} (1994), 227. (a)\\
Pardy, M.,  Phys. Lett. B {\bf 325} (1994), 517. (b)\\
Pardy, M.,  Int. J. of Theor. Phys. {\bf 34}, N. 6, (1995), 951.\\
Pardy, M.,  Phys. Rev. A  {\bf 55} No. 3, (1997), 1647. \\
Schwinger, J., Tsai W.Y. and Erber, T, Ann. Phys. (NY) {\bf 96} (1976), 303.\\
Schwinger, J., "Particles, Sources and Fields", Vol. I,
Addison-Wesley, Reading, Mass., 1970.\\
Schwinger, J., "Particles, Sources and Fields", Vol. II,
Addison-Wesley, Reading, Mass., 1973.\\
Tamm, I. E. and Frank, I. M., Dokl. Akad. Nauk USSR  {\bf 14}
(1937), 107.\\
\end{document}